\documentclass[fleqn,usenatbib]{mnras}

\usepackage{amssymb}

\usepackage{newtxtext,newtxmath}
\usepackage{accents}
\newcommand{\ubar}[1]{\underaccent{\bar}{#1}}

\usepackage[T1]{fontenc}

\DeclareRobustCommand{\VAN}[3]{#2}
\let\VANthebibliography\thebibliography
\def\thebibliography{\DeclareRobustCommand{\VAN}[3]{##3}\VANthebibliography}

\usepackage{graphicx}	
\usepackage{amsmath}	
\usepackage{amssymb}	

\title[Star formation in UNIONS post-mergers]{Star formation characteristics of CNN-identified post-mergers in the Ultraviolet Near Infrared Optical Northern Survey (UNIONS)}
\author[R. W. Bickley et al.]{Robert W. Bickley,$^{1}$\thanks{E-mail: rbickley@uvic.ca}
Sara L. Ellison,$^{1}$
David R. Patton,$^{2}$
\newauthor
Connor Bottrell,$^{3}$
Stephen Gwyn,$^{4}$
Michael J. Hudson$^{5,6,7}$
\\
$^{1}$Department of Physics and Astronomy, University of Victoria, Victoria, British Columbia V8P 1A1, Canada\\
$^{2}$Department of Physics and Astronomy, Trent University, 1600 West Bank Drive, Peterborough, ON K9L 0G2, Canada\\
$^{3}$Kavli Institute for the Physics and Mathematics of the Universe (WPI), UTIAS, University of Tokyo, Kashiwa, Chiba 277-8583, Japan\\
$^{4}$Canadian Astronomy Data Centre, NRC Herzberg, 5071 West Saanich Road, Victoria, BC, V9E 2E7, Canada\\
$^{5}$Department of Physics and Astronomy, University of Waterloo, 200 University Ave W, Waterloo, ON N2L 3G1, Canada\\
$^{6}$Waterloo Centre for Astrophysics, University of Waterloo, 200 University Ave W, Waterloo, ON N2L 3G1, Canada\\
$^{7}$Perimeter Institute for Theoretical Physics, 31 Caroline St. North, Waterloo, ON N2L 2Y5, Canada\\
}

\date{Accepted XXX. Received YYY; in original form ZZZ}

\pubyear{2021}

\begin{document}
\label{firstpage}
\pagerange{\pageref{firstpage}--\pageref{lastpage}}
\maketitle

\begin{abstract}
The importance of the post-merger epoch in galaxy evolution has been well-documented, but post-mergers are notoriously difficult to identify. While the features induced by mergers can sometimes be distinctive, they are frequently missed by visual inspection. In addition, visual classification efforts are highly inefficient because of the inherent rarity of post-mergers (\textasciitilde1\% in the low-redshift Universe), and non-parametric statistical merger selection methods do not account for the diversity of post-mergers or the environments in which they appear. To address these issues, we deploy a convolutional neural network (CNN) which has been trained and evaluated on realistic mock observations of simulated galaxies from the IllustrisTNG simulations, to galaxy images from the Canada France Imaging Survey (CFIS), which is part of the Ultraviolet Near Infrared Optical Northern Survey (UNIONS). We present the characteristics of the galaxies with the highest CNN-predicted post-merger certainties, as well as a visually confirmed subset of 699 post-mergers. We find that post-mergers with high CNN merger probabilities (p(x)>0.8) have an average star formation rate that is 0.1 dex higher than a mass- and redshift-matched control sample. The SFR enhancement is even greater in the visually confirmed post-merger sample, a factor of two higher than the control sample.
\end{abstract}




\section{Introduction}

Major mergers usher in an era of dramatic and rapid evolution for the host galaxies, simultaneously changing their dynamic, morphological, and intrinsic characteristics (\citealp{1978MNRAS.183..341W}; \citealp{1993MNRAS.262..627L}; \citealp{2008MNRAS.383...93B}; \citealp{2008ApJ...675.1095J}). N-body (e.g. \citealp{1972ApJ...178..623T}; \citealp{Conselice_2006}) and hydrodynamical (e.g. \citealp{2008MNRAS.391.1137L}) simulations are in broad agreement with observations (\citealp{2010MNRAS.401.1043D}; \citealp{2015ApJS..221...11K}; \citealp{2017MNRAS.464.4420S}) about the morphologies produced by ongoing and completed mergers, including stellar shells, streams, and bridges. Models also predict the host of ways in which the intrinsic properties of galaxies are changed by a merger (e.g. \citealp{1972ApJ...178..623T}; \citealp{2005MNRAS.361..776S}; \citealp{2008AN....329..952D}; \citealp{2019MNRAS.485.1320M}; \citealp{2020MNRAS.494.4969P}; \citealp{2020MNRAS.493.3716H}), again in agreement with observations. Central starbursts may be induced, as cold gas is funneled to the centres of galaxies by the interaction (\citealp{2008AJ....135.1877E, 2013MNRAS.435.3627E}; \citealp{2004MNRAS.355..874N}; \citealp{2012MNRAS.426..549S}; \citealp{2014MNRAS.437.2137S}; \citealp{2015HiA....16..326K}; \citealp{2019MNRAS.482L..55T}). This same cold gas may accrete onto a galaxy's super-massive black hole and trigger an active galactic nucleus (AGN; e.g. \citealp{2014MNRAS.441.1297S}, \citealp{2011MNRAS.418.2043E, 2019MNRAS.487.2491E}). Together, increased star formation activity and AGN feedback may drive strong outflows that may enrich the circum-galactic medium (\citealp{2015MNRAS.449.3263J}; \citealp{2018MNRAS.475.1160H}). Galaxies' stellar and gas kinematics are also disrupted by mergers (\citealp{1967MNRAS.136..101L}; \citealp{1977egsp.conf..401T}; \citealp{1983MNRAS.205.1009N}; \citealp{1992ApJ...400..460H}; \citealp{2003ApJ...597..893N}; \citealp{2006ApJ...645..986R}; \citealp{2009MNRAS.397.1202J}; \citealp{2014MNRAS.440L..66B}; \citealp{2018MNRAS.478.3994C}; \citealp{2018MNRAS.475.1160H}), and galaxies may quench rapidly by either ejective or preventative means after the epoch of intense star formation and AGN feedback is complete (\citealp{1988ApJ...325...74S}; \citealp{2006ApJS..163....1H}; \citealp{2014ApJ...792...84Y}; \citealp{2021MNRAS.504.1888Q}).

Any statistical effort to study typical merger-induced phenomena in observations requires a high-quality merger sample that is both pure (containing a high fraction of galaxies that are genuine mergers) and relatively complete (identifying as many of the available mergers as possible for inclusion). In studies of galaxy pairs, completeness and purity are both realistic goals due to the relative ease with which pairs can be identified visually (e.g. \citealp{2007ApJS..172..329K}; \citealp{2005ApJ...625..621B}; \citealp{1998ApJ...499..112B}; \citealp{2010MNRAS.401.1043D}) or statistically, by identifying galaxies that are close in both angular position and line-of-sight radial velocity. For large spectroscopic surveys like the Sloan Digital Sky Survey (SDSS), automated identification has proven effective to identify statistically large galaxy pair samples and study the effects of the pair (pre-coalescence) phase of the merger sequence (e.g. \citealp{2000ApJ...536..153P}; \citealp{2000ApJ...530..660B}; \citealp{2004ApJ...617L...9L}; \citealp{2005AJ....130.1516D}; \citealp{2008ApJ...681..232L}).

In order to trace the evolutionary changes experienced by galaxies along the entire merger sequence, it is also important to study post-mergers, i.e. galaxies which have fully coalesced with their interacting companion(s). Such merger remnants are no longer separated from their companions, and so are more difficult to identify statistically, even though they exhibit the morphological signatures of a recent merger. As a result, post-merger samples tend to be either quite small in number (e.g. \textasciitilde100 visually-identified post-mergers in \citealp{2013MNRAS.435.3627E}), or dubious in their purity. Once identified, however, post-mergers can subsequently be studied photometrically, spectroscopically, or in spatial detail via integral field spectroscopy (IFS) instruments (e.g. \citealp{2019MNRAS.482L..55T}; \citealp{2015A&A...582A..21B}; \citealp{2019ApJ...881..119P}, \citealp{2020arXiv200902974W}). The post-merger epoch of the merger sequence is therefore  provisionally understood. Notably, \citet{2013MNRAS.435.3627E} found that their visually-identified post-merger sample was transforming more dramatically than galaxy pair samples (at any separation) in star formation rate (SFR), metallicity, and AGN. However, the quantitative results of extant post-merger studies require further appraisal on account of small samples and potential sample inhomogeneity.

In general, post-merger identification requires imaging data of adequate volume, resolution, and depth to identify the low surface brightness features associated with post-mergers (e.g. \citealp{2019MNRAS.486..390B}), and a classification method that is highly accurate in its identification of post-mergers in the specific context of that data. The Canada France Imaging Survey (CFIS), part of the Ultraviolet Near Infrared Optical Northern Survey (UNIONS) collaboration, fulfills the first of these criteria, with \textasciitilde0.6 arcsecond seeing $r$-band imaging planned over 5,000 square degrees of the sky. Even with a conservative estimate of the post-merger incidence rate in the low-redshift universe, CFIS is expected to have imaged thousands of post-mergers (\citealp{1993MNRAS.262..627L}; \citealp{2011ApJ...742..103L}; \citealp{2012ApJ...747...34B}; \citealp{2014MNRAS.445.1157C}; \citealp{2015MNRAS.449...49R}; \citealp{2018MNRAS.480.2266M}). The survey's 5-$\sigma$ point-source depth (24.85 mag in the $r$-band for the MegaCam wide-field optical imager) is sufficient to capture the low-surface brightness features associated with mergers in the local Universe. Finally, \textasciitilde3300 square degrees of the survey overlap with the SDSS Baryon Oscillation Spectroscopic Survey (BOSS) (\citealp{2013AJ....145...10D}) footprint. Post-mergers identified in the survey will therefore be available for immediate spectroscopic characterization.

The challenge of a robust and well-adapted post-merger identification method remains, however. The task of identifying a pure and complete post-merger sample is discussed by \citet{2021MNRAS.504..372B}, in which the strengths and weaknesses of popular post-merger identification methods were evaluated on a sample of post-merger and non-post-merger galaxies from the 100-1 run of the IllustrisTNG magnetohydrodynamical simulations (\citealp{2018MNRAS.480.5113M}; \citealp{2018MNRAS.477.1206N}; \citealp{2018MNRAS.475..624N}; \citealp{2018MNRAS.475..648P}; \citealp{2018MNRAS.475..676S}; \citealp{2019ComAC...6....2N}) with added realism to mimic the properties of CFIS.  While visually-identified samples can exceed in purity, the statistical completeness (the fraction of true post-mergers correctly identified by each person) was shown to be inconsistent between individual volunteers. Traditional non-parametric statistical methods of post-merger identification, including Gini-M20, asymmetry, and shape asymmetry were shown to have limited effect in the specific context of post-merger identification. This is likely because merger-induced morphological phenomena are often faint, and can be easily be missed on account of survey artifacts, bright foreground stars, and noise. Moreover, when irregular morphologies arise in non-post-mergers, they may delude non-parametric statistical techniques and human classifiers alike.

Even for the most highly-trained visual classifier, the number of galaxies observed in CFIS presents the additional challenges of classification time and reproducibility. In a strictly visual classification exercise, inspecting the entire CFIS galaxy sample would represent an improbable task, even with months of continuous work. Further, should another member of the community be interested in re-evaluating the quality of such a sample, they would need to repeat this work for themselves. Automation is therefore a favourable approach (e.g. \citealp{Conselice_2003}; \citealp{2004AJ....128..163L}; \citealp{2016MNRAS.456.3032P}; \citealp{2019MNRAS.483.4140R}; \citealp{2019ApJ...872...76N}), and degrees of success in identifying larger and more reproducible post-merger samples have already been achieved.

\citet{2021MNRAS.504..372B} argue that convolutional neural networks (CNNs), which have been applied to a range of problems within and outside of astronomy (e.g. \citealp{2015ApJS..221....8H}; \citealp{2018MNRAS.476.3661D}; \citealp{2019ApJS..243...17J}; \citealp{2019MNRAS.484...93D}; \citealp{2019ApJ...876...82N}; \citealp{2019MNRAS.489.1859H}; \citealp{2020ApJS..248...20H}) are the most promising for rapid identification of a statistically large and high-quality post-merger sample. The only classification biases held by a newly-trained CNN are those derived from the training data. Further, the learned neuron weights of the resulting network are portable, and can be re-applied by the wider community to reproduce and further study classification efforts. Likely because of these strengths, astronomers interested in mergers have already applied CNNs successfully to the task of merger identification in imaging surveys (e.g. \citealp{2018MNRAS.479..415A}; \citealp{2019MNRAS.483.2968W}, \citealp{2019A&A...626A..49P}, \citealp{2020ApJ...895..115F}, \citealp{2020arXiv200902974W}) as well as in stellar velocity fields (e.g. \citealp{2022MNRAS.511..100B}).

The success of any effort to applying a CNN to an astrophysical problem is limited by two main factors: 1) the quality of the training data, including its size, and similarity to the astrophysical data to which it will ultimately be applied, and 2) the strength of the physical connection between the network's input (in this case, galaxy images) and the desired output (a prediction of post-merger status). \citet{2019MNRAS.490.5390B}, \citet{2019MNRAS.489.1859H}, and \citet{2021arXiv211100961C} have shown that observational realism is of paramount importance for the success of a CNN in the context of morphological classification. Naturally, it is critical that the data to which a CNN will ultimately be applied (in this case, CFIS imaging) is of similar composition as the training data. In a broader sense, it is important that the training post-mergers and non-post-mergers represent the true range of environments, merger stages, and circumstances in which galaxies in the Universe might be found. \citet{2021MNRAS.504..372B} address this by training a CNN on carefully selected, equally sized post-merger and non-post-merger samples from IllustrisTNG 100-1 that are matched on environmental parameters and mass, and assigned redshifts, observational point-spread function (PSF) values, and sky backgrounds taken from the actual survey. \citet{2021MNRAS.504..372B} generate mock observations from the selected parameters for each galaxy assuming a Planck15 (\citealp{2016A&A...594A..13P}) cosmology.

Using simulated galaxies in training also resolves the second aforementioned limitation. Cosmological simulations allow for unambiguous access to the merger status of the galaxies they contain, ensuring the strongest possible connection between the input imaging data and the merger status of the training galaxies. When applying the CNN to real CFIS galaxies, however, this connection is thrown into doubt - if IllustrisTNG mergers do not share their morphological characteristics with mergers in the Universe, the success of the CNN will be undermined. Thankfully, this does not appear to be the case: \citet{2019MNRAS.483.4140R} reported good agreement (\textasciitilde1-$\sigma$ for all parameters studied) between the optical morphologies of IllustrisTNG galaxies processed with \textsc{skirt} radiative transfer\footnote{skirt.ugent.be} (\citealp{2011ascl.soft09003B}; \citealp{2015A&C.....9...20C}), and real galaxies observed by Pan-STARRS (\citealp{2016arXiv161205560C}). \citet{2019MNRAS.487.5416T}, \citet{2019MNRAS.489.1859H}, \citet{2020ApJ...895..139D}, and \citet{2021MNRAS.501.4359Z} also find that similar feature spaces are shared by IllustrisTNG and real galaxies.

\citet{2021MNRAS.504..372B} found that the network performed well on simulation galaxies set aside for testing, both in a balanced sample and in a mock survey composed of over 300,000 simulated galaxies processed with CFIS realism (see Section~\ref{Methods} for details of this performance). Merger status in the work was given by the CNN as a p(x) prediction, analogous to post-merger certainty, on a scale of 0 to 1. After validating the network in this context, it was argued that a high CNN decision threshold (a cut in p(x), ruling out many galaxies with less certain post-merger statuses) could be used to significantly increase the purity of post-mergers in the out-falling sample. After distilling the sample in this way, its constituent galaxy images could be inspected visually, and questionable (or obviously inaccurate) post-merger predictions could be discarded much more efficiently.

In this work, we revisit the preparation and characteristics of the simulation-trained CNN, and prepare CFIS images of galaxies for processing by the network (Section~\ref{Methods}). We next apply the CNN to the CFIS images so prepared and validate its predictions statistically (Section~\ref{CNN-identified post-mergers}), perform rigorous visual inspection, and present a new sample of 699 visually confirmed post-merger galaxies (Section~\ref{The hybrid classification post-merger sample}). Finally, we study the effect of modulating the CNN decision threshold on the star formation characteristics of the CNN-predicted and visually inspected post-merger samples (Section~\ref{Modulating the decision threshold}), and revisit the \citet{2013MNRAS.435.3627E} characterization of star formation enhancement ($\Delta$SFR) along the merger sequence with a post-merger sample of novel statistical size and purity (Section~\ref{Star formation in the merger sequence}).

\section{Methods}
\label{Methods}

\subsection{Neural network preparation}

\citet{2021MNRAS.504..372B} detailed the creation of a CNN trained on balanced samples of post-merger and non-post-merger galaxies from the 100-1 run of the IllustrisTNG cosmological magnetohydrodynamical simulations. Post-mergers were identified using the subhalo merger trees created by \textsc{Sublink} (\citealp{2015MNRAS.449...49R}), following the methodology of \citet{2020MNRAS.493.3716H}. Merger remnants in the post-merger training set were required to have stellar mass ratios of $\mu \geq 0.1$, occur at $z \leq 1$ (snapshots 50-99 of the simulation), and fall within the stellar mass range $\mathrm{10^{10}-10^{12}}$ \(\textup{M}_\odot\). In order to limit the search to recent post-mergers, \citet{2021MNRAS.504..372B} only included galaxies that had completed a merger during the most recent simulation snapshot, i.e. within \textasciitilde150 Myr. A sample of non-post-merger galaxies, which had not undergone a merger in at least 2 Gyr, were matched to the post-merger sample on stellar mass, simulation snapshot number, and a suite of environmental parameters. After stellar mass maps were extracted from the simulations and assigned an intrinsic brightness from the \citet{2019ComAC...6....2N} stellar photometrics tables, they were processed with RealSimCFIS, a custom version of the observational realism code RealSim\footnote{github.com/cbottrell/RealSim} originally developed for the SDSS and detailed in \citet{2019MNRAS.490.5390B}. RealSimCFIS first dims and rebins each TNG galaxy according to an observational redshift chosen at random from the distribution of real CFIS galaxies. An observational point-spread function is applied next. Finally, real skies from CFIS are added to each image in order to maximize the verisimilitude of the synthetic images and to prepare the network for the diversity of observational effects that it will encounter in the real survey.

The model's architecture, loosely based on AlexNet (\citealp{NIPS2012_c399862d}) and refined through a manual hyperparameter search, is given in Table \ref{arx-table}. After training and testing the network, \citet{2021MNRAS.504..372B} find that the model successfully identifies \textasciitilde89\% of post-mergers and non-post-mergers in a mock survey composed of unseen synthetic galaxy images produced with the same pipeline as the training data. Moreover, the network's ability to identify post-mergers remained stable across a range of galaxy environments, observational redshifts, stellar masses, mass ratios, and gas fractions. \citet{2021MNRAS.504..372B} found that the network had some difficulty in distinguishing between post-mergers and close pre-coalescence pairs, but postulated that galaxy pairs are rare, and therefore might not present a significant obstacle. More pressing, however, is the inherent rarity of the post-merger class. Because post-mergers only constitute \textasciitilde1\% or less of galaxies in the low-$z$ Universe (\citealp{2000ApJ...536..153P}), any classification tool designed to identify them must have incredibly high accuracy in order to produce a sample that is reasonably pure (see Figure 1 in \citealp{2022MNRAS.511..100B}). In \citet{2021MNRAS.504..372B}, two courses of action are recommended to remedy this issue:

(1) The convolutional network detailed in \citet{2021MNRAS.504..372B} and applied in this work returns predictions of post-merger status on a sliding scale between 0 (very unlikely to be a post-merger) and 1 (highly likely to be a post-merger). If post-mergers and non-post-mergers were equally abundant in the universe, a decision threshold of 0.5 might be a natural choice to discriminate between the two classes. However, \citet{2021MNRAS.504..372B} show that one can exchange completeness (the fraction of genuine post-mergers that are recovered) for purity (the fraction of predicted post-mergers that are true post-mergers) by increasing the value of this decision threshold. It was also shown that post-merger populations identified using a higher decision threshold exhibited $\Delta$SFRs that were better aligned with the ground truth of the simulation.

\begin{figure}
\includegraphics[width=\columnwidth]{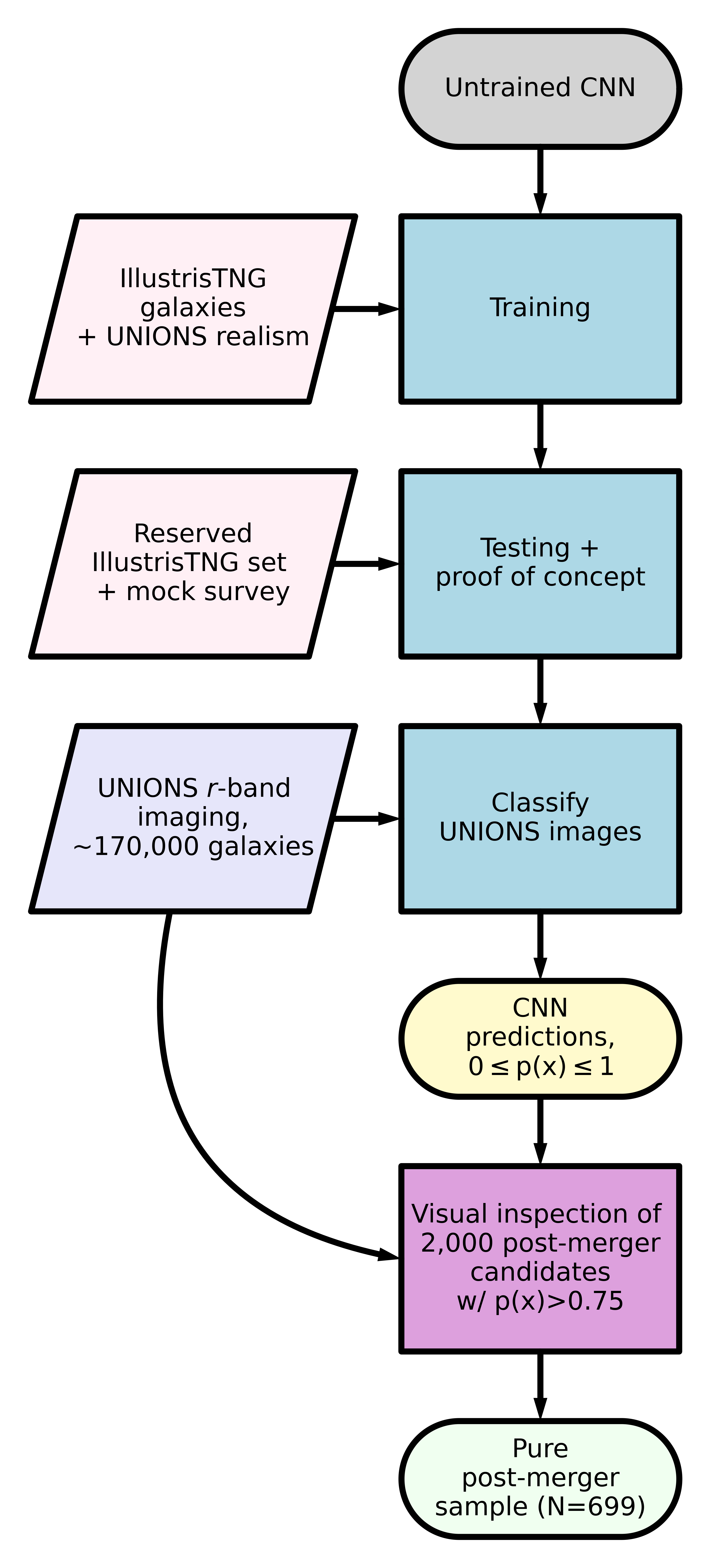}
\caption{A flowchart showing the inputs and outputs (ovals), procedures (boxes), and data sets (parallelograms) described in \citet{2021MNRAS.504..372B} and this work. Our CNN is trained and tested on example post-merger and non-post-merger galaxies from the IllustrisTNG simulations before it is used to obtain post-merger classifications on real galaxies in $r$-band UNIONS imaging. The classifications are then inspected visually, and a sample of post-mergers is confirmed and presented.}
\label{fig:SFR-flowchart}
\end{figure}

(2) Expert visual classifications have long been considered the gold standard in post-merger identification. \citet{2021MNRAS.504..372B} suggest that a hybrid approach, in which a small subset of galaxies identified as strong post-merger candidates are subsequently inspected by a trained person, may be the optimal way forward. Humans, for example, may be able to alleviate the CNN's difficulty with close galaxy pairs, as double nuclei are often visually distinct. Combining automated CNN and visual classification within the framework of a hybrid experiment should therefore efficiently produce a sample that is both highly pure and greater in size than would be possible for an unaided human classifier.

The flowchart in Figure~\ref{fig:SFR-flowchart} illustrates how a CNN can be trained on simulated data, applied to observed galaxies, and combined with visual classifications to produce a large, homogeneously selected, and highly pure post-merger sample. It is this combination of methods that we ultimately use to identify the post-merger sample described in Section~\ref{The hybrid classification post-merger sample}.

\begin{table}
\begin{center}
\begin{tabular}{ |c|c|c|c| } 
\hline
Layer Type & \# Parameters & Output Shape\\
\hline
\hline
Input & 0 & (138,138,1) \\ 
\hline
\begin{tabular}{@{}@{}c@{}}Convolution \\ 32 Filters \\ Kernel (7,7)\end{tabular}  & 1600 & (138, 138, 32) \\ 
\hline
Max Pooling (2,2) & 0 & (69, 69, 32) \\ 
\hline
Dropout 25\% & 0 & (69, 69, 32)\\ 
\hline
\begin{tabular}{@{}@{}c@{}}Convolution \\ 64 filters \\ Kernel (7,7)\end{tabular} & 100416 & (69, 69, 64) \\ 
\hline
Max Pooling (2,2) & 0 & (34, 34, 64) \\ 
\hline
Dropout 20\% & 0 & (34, 34, 64) \\ 
\hline
Batch Normalization & 256 & (34, 34, 64) \\
\hline
\begin{tabular}{@{}@{}c@{}}Convolution \\ 128 filters \\ Kernel (7,7)\end{tabular} & 401536 & (34, 34, 128) \\ 
\hline
Max Pooling (2,2) & 0 & (17, 17, 128) \\ 
\hline
Dropout 20\% & 0 & (17, 17, 128) \\ 
\hline
\begin{tabular}{@{}@{}c@{}}Convolution \\ 128 filters \\ Kernel (7,7)\end{tabular} & 802944 & (17, 17, 128) \\ 
\hline
Max Pooling (2,2) & 0 & (8, 8, 128) \\ 
\hline
Dropout 20\% & 0 & (8, 8, 128) \\ 
\hline
\hline
Flatten & 0 & (8192) \\
\hline
Dense & 4194816 & (512) \\
\hline
Dropout 25\% & 0 & (512) \\
\hline
Dense & 65664 & (128) \\
\hline
Dropout 25\% & 0 & (128) \\
\hline
Activation, Sigmoid & 129 & (1) \\
\hline
\hline
\end{tabular}
\end{center}
\caption{The CNN architecture used in this work. Each layer begins with the stock Keras layer of the same name, with any specified hyperparameters detailed in the Layer Type column. The \# Parameters column shows the number of trainable network parameters belonging to each layer.}
\label{arx-table}
\end{table}

\subsection{Multi-survey data}

High quality imaging resolution and depth over a large sky area are critical to our post-merger identification effort. The UNIONS collaboration is a new consortium of wide field imaging surveys of the northern hemisphere and represents an excellent opportunity for merger searches. UNIONS consists of CFIS conducted at the 3.6-meter CFHT on Maunakea, members of the Pan-STARRS team, and the Wide Imaging with Subaru HyperSuprimeCam of the Euclid Sky (WISHES) team. CFHT/CFIS is obtaining deep $u$ and $r$-bands; PanSTARRS is obtaining deep $i$ and moderate-deep $z$-band imaging, and Subaru is obtaining deep $g$-band (Waterloo-Hawaii IfA $g$-band Survey, WHIGS) and $z$-band (Wide Imaging with Subaru HSC of the Euclid Sky, WISHES) imaging. These independent efforts are directed, in part, to securing optical imaging to complement the Euclid space mission, although UNIONS is a separate consortium aimed at maximizing the science return of these large and deep surveys of the northern skies. \citet{2021MNRAS.504..372B} used only the $r$-band imaging in creating synthetic images, and the CNN predictions used in this work are made exclusively on $r$-band imaging.

The observing pattern employed by CFIS uses three single-exposure visits with field-of-view (FOV) offsets in between for optimal astrometric and photometric calibration with respect to observing conditions. This also ensures that the entire survey footprint will be visited for at least two exposures. After raw images are collected by CFHT, they are detrended (i.e. the bias is removed and the images are flat-fielded using night sky flats) with the software package MegaPipe (\citealp{2008PASP..120..212G}). The images are next astrometrically calibrated using Gaia DR2 (\citealp{2016A&A...595A...1G,2018A&A...616A...1G}) as a reference frame. Pan-STARRS 3$\pi$ $r$-band photometry (\citealp{2016AAS...22732407C}) is used to generate a run-by-run differential calibration across the MegaCam mosaic, and an image-by-image absolute calibration. Finally, the individual images are stacked onto an evenly spaced grid of 0.5-degree-square tiles.

The CNN was trained and previously evaluated on synthetic galaxy images that had been cropped to a physical scale of 100 kpc on a side, upscaled or downscaled to 138$\times$138 pixels (corresponding to the natural pixel resolution of a galaxy imaged by CFIS at the median redshift of the sample), and normalized on a linear scale so that the brightest pixel had a value of 1, and the faintest had a value of 0. In order to prepare galaxies imaged in CFIS data release 2 (DR2) for evaluation in the same way, we therefore require access to an accurate (i.e. spectroscopic) redshift measurement for each object. To this end, we limit our post-merger search to galaxies with $z$<0.5 in the overlap between CFIS DR2 and SDSS data release 7 (DR7), and match objects in the two catalogs with a 2-arcsecond tolerance. The resulting sample contains 168,597 galaxies. We use the SDSS-derived spectroscopic redshift to produce CFIS cutouts with the same field of view, pixel size, and normalization characteristics of the images used for training and evaluation in \citet{2021MNRAS.504..372B}. The availability of SDSS spectra for post-merger galaxies will also facilitate subsequent spectroscopic characterization of the galaxies' properties (see Section~\ref{Star formation in post-mergers}).

\section{The CFIS post-merger sample}
\label{The CFIS post-merger sample}

After preparing our CNN classifier and CFIS galaxy images, the CNN makes a prediction of post-merger status on each image. In this section, we seek to statistically appraise the quality of the CNN's post-merger predictions in an observational context. We then present a catalog of CNN-predicted post-mergers that also passed rigorous visual inspection by the authors.

\subsection{CNN-identified post-mergers}
\label{CNN-identified post-mergers}

The true merger status of galaxies in the Universe is unknown to us as observers. Therefore, in analyzing the results of a post-merger identification effort, robust experimental validation is also elusive, and sample impurity is expected. Nevertheless, we can check for systematic issues by inspecting the global statistical characteristics of the out-falling galaxy samples. Because post-mergers are rare, for example, if a CNN (or other merger identification tool) found that a survey was comprised mostly of post-mergers, there would be cause for alarm. We can also check for biases by comparing the physical characteristics of the CNN-predicted post-mergers to those of the parent sample (CFIS galaxies with available SDSS spectra).

\begin{figure}
\includegraphics[width=\columnwidth]{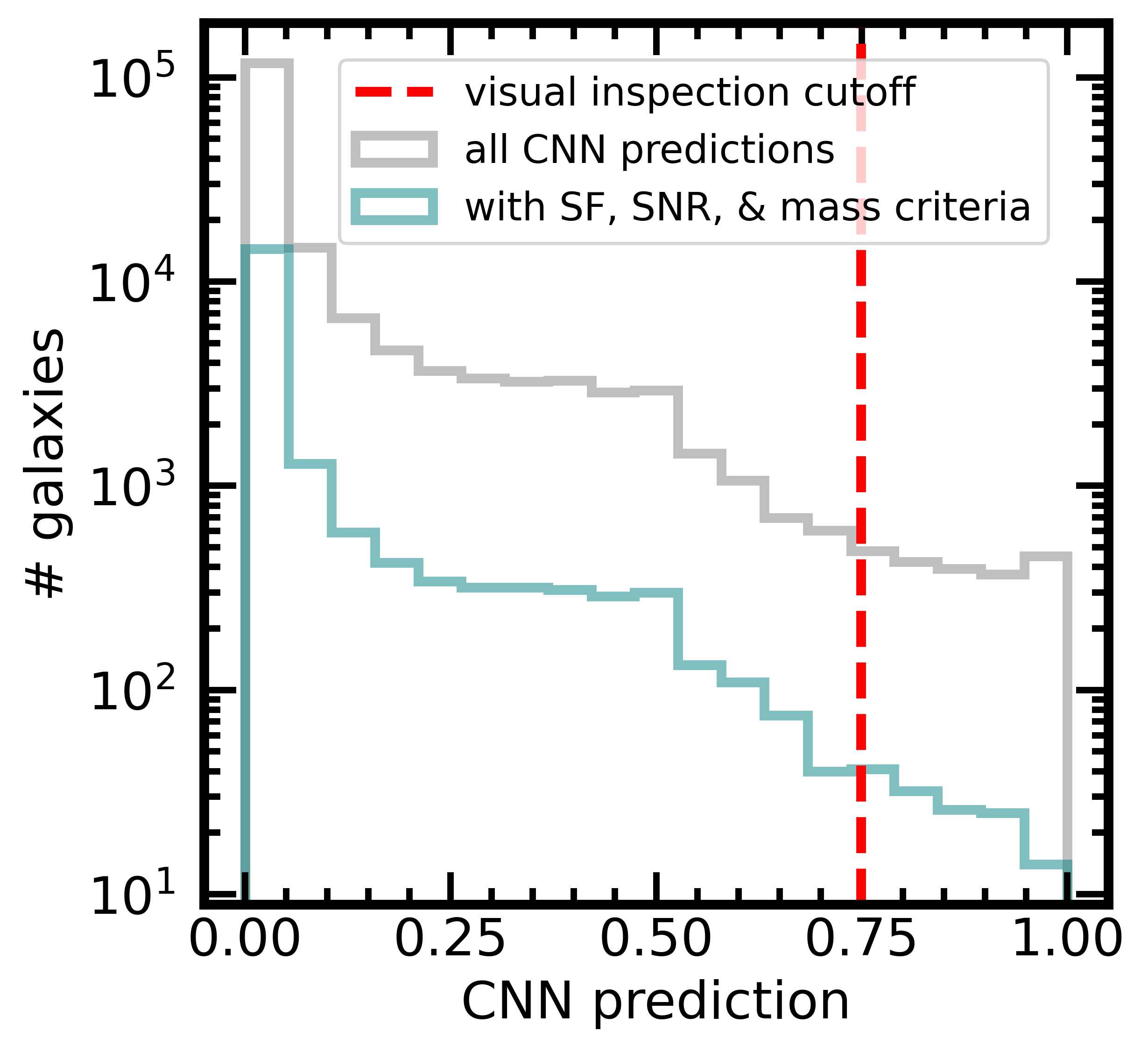}
\caption{The evaluated CFIS galaxies in bins of CNN p(x) (grey), the subset of those galaxies meeting the star formation, signal to noise, and mass criteria for our star formation enhancement experiment (Section~\ref{Star formation in post-mergers}, teal), and the p(x) threshold (red dashed line) above which the galaxies were visually inspected by the authors.}
\label{fig:thresh-hist}
\end{figure}

Figure~\ref{fig:thresh-hist}, which shows two histograms of the predictions made by the CNN, addresses the systematic approach to validation. The grey histogram shows the log-binned CNN predictions between 0 and 1 assigned to CFIS galaxies with SDSS spectra (hereafter, the "parent sample"). As in the mock survey detailed in \citet{2021MNRAS.504..372B}, the CNN assigns non-post-merger predictions to the vast majority of galaxies when we use the "default" decision threshold at 0.5 to differentiate between the two classes. Naturally, more strict (higher) decision thresholds can be used to force even greater selectivity: \citet{2021MNRAS.504..372B} demonstrated that in an unstudied test set of synthetic galaxy images, the network was very well-calibrated, meaning that the proportion of true post-mergers in bins of p(x) scaled with the value of p(x). Consequently, if CFIS galaxies are sufficiently visually similar to the training galaxies from IllustrisTNG 100-1, the CNN's numerical post-merger prediction on each CFIS galaxy can be interpreted as probability-like.  We therefore count galaxies with p(x)>0.75 as highly likely to be post-mergers in this subsection, and flag them for visual inspection in Section~\ref{The hybrid classification post-merger sample}. Although the CNN's p(x) axis represents opposing gradients of completeness (the number of true post-mergers included above some value) and purity (the fraction of galaxies above some value that are true post-mergers), we choose p(x)>0.75 as a convenient criterion for "highly likely" because \textasciitilde2,000 galaxies lie above this value (represented in Figure~\ref{fig:thresh-hist} by the dashed red line), and 2,000 galaxies can be feasibly inspected by eye for the purposes of this experiment. For reference, the galaxies meeting the spectroscopic signal to noise (S/N), SFR, and mass criteria for our star formation enhancement experiment (detailed in Section~\ref{Star formation in post-mergers}) are shown on Figure~\ref{fig:thresh-hist} as well.

\begin{figure*}
\includegraphics[width=\textwidth]{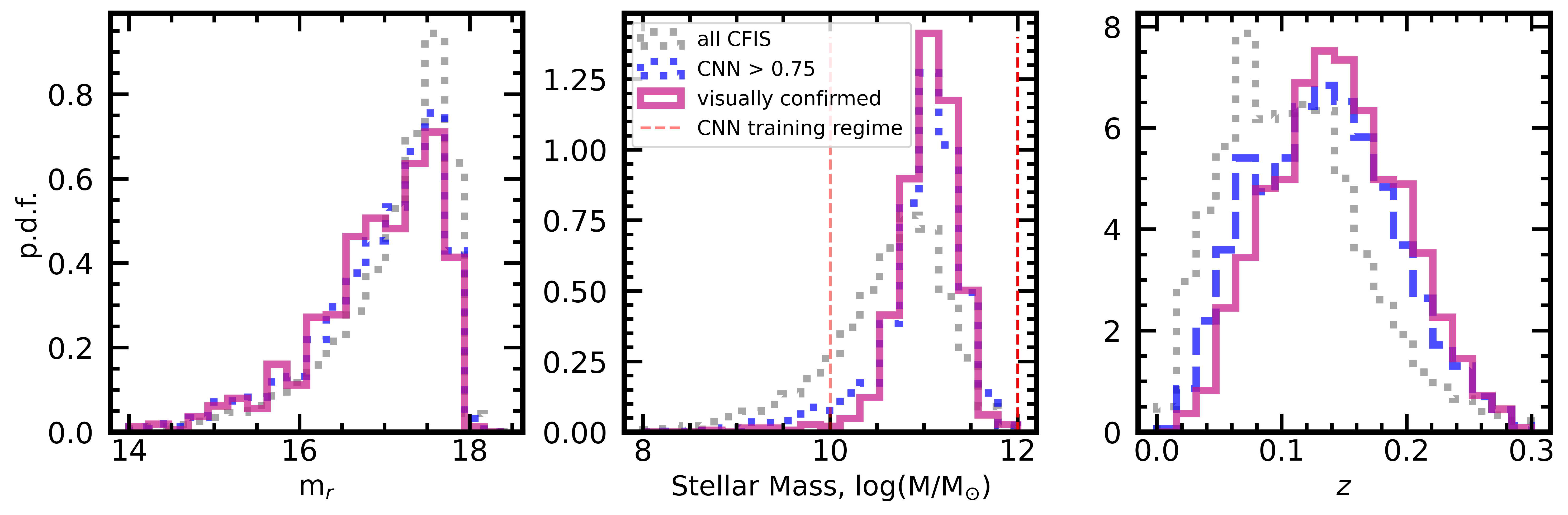}
\caption{The $r$-band Petrosian magnitudes (left), stellar masses from \citet{2014ApJS..210....3M} (centre), and redshifts (right) of the parent sample of CFIS galaxies with SDSS spectra (grey dashed histogram), the CNN parent sample with predictions >0.75 (blue dashed histogram), and the 699 visually confirmed post-mergers (violet histogram, see Section~\ref{The hybrid classification post-merger sample}). All three histograms are normalized density distributions, for simplicity of comparison. The three samples are fairly consistent with one another in r-band magnitude, suggesting that the CNN's training on images with a realistic range of imaging signal to noise was broadly effective. There is a deficit of very faint galaxies (see m$_{r}$>17.5) in the post-merger sample (left panel). Individual inspection of these galaxies reveals that they have relatively low stellar masses, and are less bright at a given distance than their more massive counterparts. The CNN's preference for more massive galaxies (centre panel) is striking, even within the mass range on which the network was trained. Because more massive galaxies are brighter relative to the background, they are more likely to exhibit bright merger signatures if they have undergone a merger. This bias is echoed in the redshift distribution (right panel), since the volume-limited nature of the survey means that very massive galaxies are more likely to appear at higher redshift. The cause of this effect is explored in Appendix~\ref{appA}.}
\label{fig:confirmed-stats}
\end{figure*}

Figure~\ref{fig:confirmed-stats} compares the physical (stellar mass and redshift) and observational ($r$-band magnitude) characteristics of the CNN-predicted post-merger sample (blue histograms) to those of the parent sample (grey histograms). In training, the CNN was shown images of a diverse population of simulated post-merger galaxies, inserted into a realistic range of CFIS galaxy redshifts, survey skies, and observational seeing in order to prepare it to encounter galaxies with similarly varied characteristics in the observations. These efforts were broadly successful; post-mergers were found across the entire observational range of brightness, stellar mass, and redshift.

Still, the predicted post-merger sample is not bias-free. In \citet{2021MNRAS.504..372B}, the network recovered \textasciitilde89\% of the post-merger galaxies in a mock survey, designed to mimic the characteristics of the observational sample it would eventually classify in this work. We therefore expect that a non-zero number of true post-merger galaxies in CFIS were given negative (<0.5) classifications by the CNN, and were excluded. Still more may have been excluded from consideration between 0.5<p(x)<0.75, where we certainly expect a small fraction of post-mergers to lurk. Because images with p(x)<0.75 were not inspected, it is difficult to speculate on the cause of their exclusion from the post-merger sample, but it is reasonable to assume that their visual disturbances were not strong enough, due to low surface brightness or low mass. Conceivably, a true post-merger might be obstructed by a UNIONS imaging artifact such that its merger status could not be determined. The statistics of the CNN-predicted post-mergers (blue histograms, Figure~\ref{fig:confirmed-stats}) also reveal that the CNN itself developed biases during its training. While the network correctly identifies a number of post-mergers at very faint apparent magnitudes, there is a deficit in the CNN-identified and visually confirmed samples relative to the parent sample. This bias is likely a natural consequence of the fact that tidal features in brighter post-merger galaxies are more visually distinct (i.e. brighter relative to the background).

This brightness preference manifests most distinctly in the stellar mass statistics of the three samples. At a given redshift, a galaxy with a higher stellar mass is very likely to shine brighter than a lower-mass galaxy. As a result, the CNN population is skewed towards higher masses, peaking near $\mathrm{10^{11}}$ \(\textup{M}_\odot\) in the middle of the network's training regime ($\mathrm{10^{10}-10^{12}}$ \(\textup{M}_\odot\)). The parent sample is volume-limited at high stellar masses, and so its mass bias causes it to prefer galaxies at a higher redshift on average as well. An argument for the brightness-motivated mass bias producing the redshift bias, rather than the other way around, is given in Appendix~\ref{appA}.

In spite of these effects, the CNN's classifications are extremely helpful as a first step in our efforts to distill a post-merger sample of novel size and purity (i.e. to produce a small sample of high post-merger concentration from a larger sample in which post-mergers are very rare). During training, the weights assigned to the CNN's neurons are determined iteratively as the network studies subsets ("batches" in neural network training parlance) of the training data. The final weights are chosen based on the moment in the network's training history when it was most successful in classifying post-mergers and control galaxies in the simulated data. Consequently, the network's learned biases may be useful in accomplishing the goal of correctly labeling as many galaxies as possible. The network's preference for brighter and more massive galaxies may be a natural consequence of the fact that the merger statuses of galaxies whose tidal features blend into the background are indeed more ambiguous.

\subsection{The hybrid classification post-merger sample}
\label{The hybrid classification post-merger sample}

\begin{figure}
\includegraphics[width=\columnwidth]{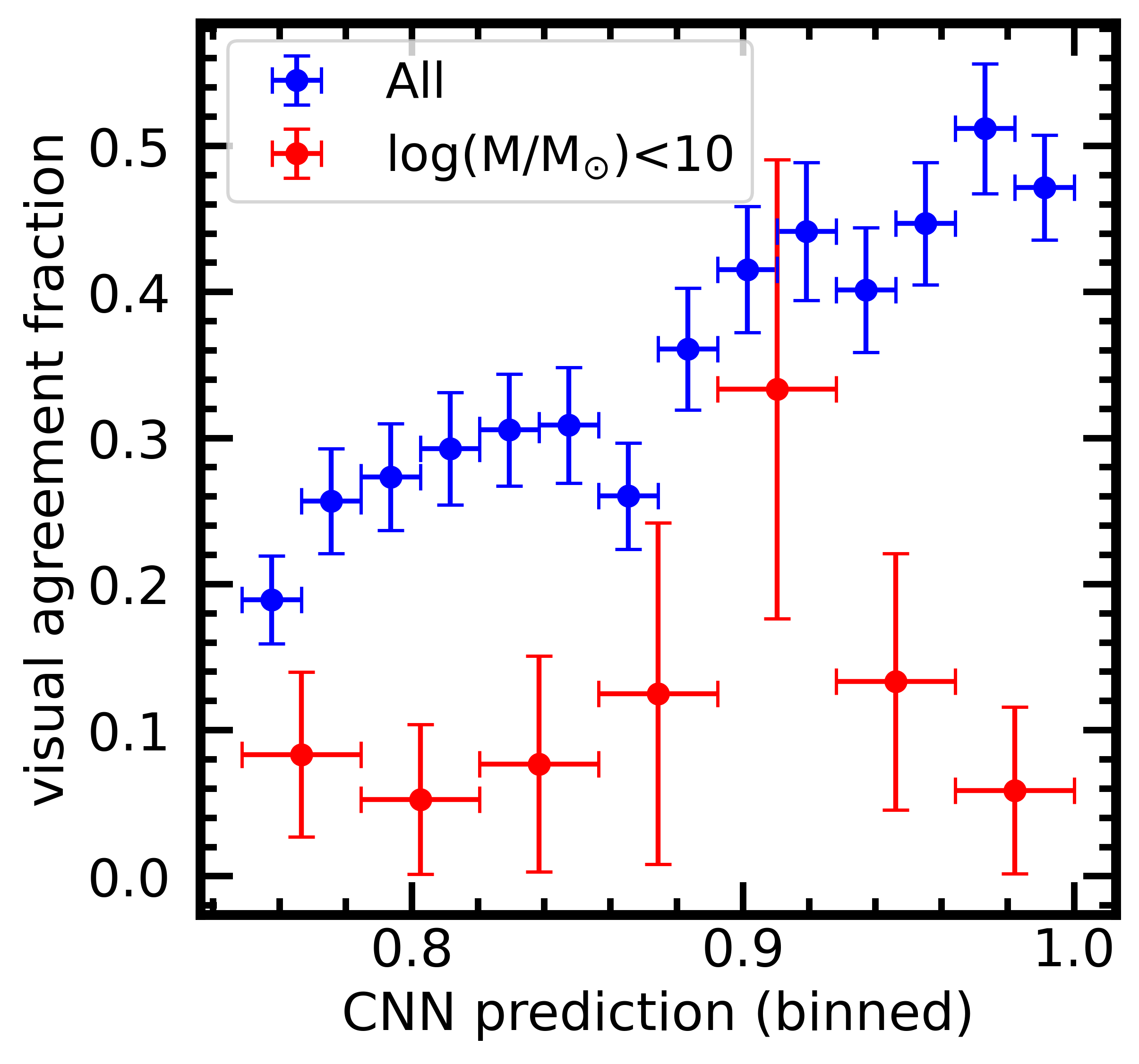}
\caption{The fraction of galaxies labeled as post-mergers by the authors in bins of CNN p(x) (blue), and the same for the subset of CNN-predicted post-mergers with masses < $\mathrm{10^{10}}$ \(\textup{M}_\odot\) (red). In general, galaxies given very high p(x) values by the CNN are the most likely to have been assigned post-merger labels by the authors. That these quantities trend positively implies that there is a meaningful connection between the criteria for post-merger status for visual inspection and the CNN. The agreement fraction is consistently low for the subset with masses < $\mathrm{10^{10}}$ \(\textup{M}_\odot\), suggesting that the CNN has a higher error rate outside of its mass training regime ($\mathrm{10^{10}-10^{12}}$ \(\textup{M}_\odot\)). Vertical error bars are the binomial error in each bin given by  $\sqrt{f*(1-f)/N}$ where f is the agreement fraction, and N is the number of galaxies in the bin. Horizontal errors are the bin widths.}
\label{fig:thresh-v-visagree}
\end{figure}

After being classified by the CNN, 2,000 galaxies given high post-merger p(x) predictions (those belonging to the grey histogram to the right of the red dashed line at \textasciitilde0.75 on Figure~\ref{fig:thresh-hist}) were in turn inspected by the author RWB, employing a purity-motivated classification philosophy in which galaxies whose post-merger statuses were in doubt were rejected. The visual classifications were performed on 100$\times$100 kpc CFIS monochromatic $r$-band image cutouts, though the cutouts were provided at their original resolution, since human visual classifications do not require a single pixel size. In many cases, SDSS full-colour images of the wider field surrounding the target galaxy were also inspected in order to verify post-merger status specifically (as opposed to a pair-phase interaction). In order to pass inspection, galaxies were required to be fully coalesced, with only one discernible nucleus. They were also required to have visible signatures of a recent merger: stellar streams, shells, and/or rings, often marked by extended and asymmetrical morphologies. Conversely, galaxies were ruled out if they were not fully coalesced, appeared to belong to a merging or interacting pair (either in CFIS or by inspection of the galaxy's wider environment via SDSS), or did not have noticeable post-merger morphological characteristics at any contrast setting. Although the CNN was trained on images that included realistic and proportionally-representative CFIS survey artifacts, occasionally a predicted post-merger would be obscured by a CFIS artifact in such a way that its merger status was indiscernible; these images were also excluded. After this first visual inspection, the systems given positive post-merger labels were inspected jointly by the authors RWB, SLE, and DRP until a consensus label was chosen for each galaxy. If one or more of the authors felt skeptical about a given galaxy's classification as a post-merger during this review, it was removed from the sample. Around 65\% of the galaxies were ultimately removed from the CNN-predicted post-merger sample with p(x)>0.75, but 699 galaxies were unanimously confirmed as post-mergers by the authors after visual inspection (hereafter, the "visually confirmed" sample). While the visually confirmed sample is certainly not complete, we believe it to be highly pure.

The results of the visual classification exercise are summarized in Figure~\ref{fig:thresh-v-visagree}. In a series of bins between the lowest (p(x)\textasciitilde0.75) and highest (p(x)=1) CNN predictions inspected by the authors, we can plot the fraction of galaxies for which the visual labels agreed with the network's positive post-merger prediction. The figure reveals a clear positive trend (blue series), suggesting that common criteria are considered by both the CNN and the trained human eye. Even for very high p(x), the agreement fraction is only \textasciitilde50\%, highlighting the need for visual inspection of the CNN-identified sample in order to increase purity. Globally, the two classification systems agreed that more than a third of the 2,000 galaxies inspected were indeed post-mergers. However, the visual classification exercise suggests that the network was less successful in identifying post-mergers outside of its mass training regime in IllustrisTNG, between $\mathrm{10^{10}-10^{12}}$ \(\textup{M}_\odot\). This mass range was selected to ensure that the simulated training galaxies were well-resolved, with at least \textasciitilde$10^{4}$ star particles. Perhaps due to this deliberate omission in training, \textasciitilde90\% of galaxies with masses < $\mathrm{10^{10}}$ \(\textup{M}_\odot\) over the entire range of CNN p(x) were not confirmed as post-mergers. Since the success of a CNN is limited by the quality and range of the training data, it is unsurprising that the network does not make accurate classifications when it is forced to extrapolate in this way. Conversely, it is not a detriment to the purity of the final post-merger sample that galaxies with masses < $\mathrm{10^{10}}$ \(\textup{M}_\odot\) were included in the observational study, since any correctly classified post-mergers below this mass threshold underwent the same rigorous visual inspection as their higher-mass counterparts in order to be included.

Galaxies were most commonly rejected by the authors due to the presence of close double nuclei that only revealed themselves under inspection with a high-contrast setting. A number of flocculent barred-spiral galaxies with a prominent ring (of approximate Hubble type SBc) were also found to contaminate the CNN-predicted sample. Less commonly, galaxies were rejected due to an obstructive survey artifact, an imaged interacting pair in either the CFIS cutout or wider SDSS imaging, or a lack of sufficiently strong merger-induced features. If the visual classifications are taken as true, we can quantitatively assess the performance of the CNN as a tool for post-merger sample distillation. If post-mergers constitute \textasciitilde0.55\% of the galaxies in the low-redshift Universe (\citealp{2000ApJ...536..153P}), the CNN-identified sample of galaxies with p(x)>0.75 is \textasciitilde64 times as post-merger-rich as the original, unprocessed galaxy sample.\footnote{0.55\% is the estimate of the merger remnant fraction over a single merger timescale, while many merger timescales have elapsed since $z$=0.5. In practice, the true post-merger fraction in our sample may therefore be higher.} Pre-filtering by p(x) is therefore a highly efficient first step in post-merger identification.

That nearly two thirds of the post-mergers with CNN p(x)>0.75 are eliminated after visual inspection suggests that the CNN produces a significant number of false positives (non-post-merger galaxies erroneously classified as post-mergers). Having completed visual classifications of all galaxies above this p(x) cut, we can speak qualitatively to the potential reasons for suspected false-positive classifications. Most commonly, the network appeared to select galaxies with one or more characteristics that mimicked the appearance of genuine post-mergers. For example, a number of interacting galaxies with extended tidal features were selected by the CNN. Such mistakes could arguably be trained away, by (for example) combining the original CNN with another one designed to identify galaxy pairs. Within the context of this experiment, we argue that such efforts would have limited utility due to the relatively small number of predicted post-mergers. In addition, if the pair classification CNN were any less than 100\% accurate in practice, there would be post-merger loss associated with network combination.

Less commonly, the CNN predicted p(x)>0.75 for galaxies of unconvincing post-merger status. While the practical origin of these predictions is unknown, it is possible that a number of post-mergers exhibiting weaker merger characteristics were shown to the network during training, in spite of our prescribed selection of major mergers (with stellar mass ratios of at least 1:10) for training in IllustrisTNG100-1. Indeed, the orbital parameters for a given galaxy merger can produce a range of visual strengths even at a fixed mass ratio. Any biases imposed due to unconvincing strength are therefore of strictly visual origin.

Statistically, the visually confirmed sample is a reasonably consistent subset of the CNN-identified sample (violet histograms in Figure~\ref{fig:confirmed-stats}), but biases were necessarily imposed by the visual classification exercise. Disturbed systems with a plausibly interacting companion were always removed from the post-merger sample, even though the presence of a companion does not preclude the possibility of post-merger status. As a result, the visually confirmed post-mergers exist in relative isolation compared to the parent sample. The visually confirmed sample is also likely biased towards more major and disruptive mergers, as a number of post-mergers were excluded due to the ambiguous (or unconvincing) strength of their tidal features. While the sample of 699 post-mergers is certainly not free of biases, we argue that it is likely representative of the remnants of low-redshift major mergers that happen to be bright on the sky. Moreover, the post-merger status of every galaxy in the visually confirmed sample is firmly defensible. This final post-merger sample is partially detailed in Table~\ref{pm-table} and available for download in full through MNRAS.

\begin{table*}
\begin{center}
\begin{tabular}{ |c|c|c|c|c|c|c|c| } 
\hline
SDSS Object ID & RA & DEC  & $z$ & Stellar Mass, log(M/\(\textup{M}_\odot\)) & \ SFR,  log(M/\(\textup{M}_\odot\)/yr) & m$_{r}$ & CNN prediction \\
\hline
\hline
587725469589831827 & 118.57040185 & 40.12469876 & 0.0688 & 11.15 & $-0.69$ & 15.32 & 0.868 \\
587725469593501956 & 125.45850409 & 47.01624984 & 0.0748 & 11.01 & $-0.62$ & 15.84 & 0.897 \\
587725471203131595 & 122.13330213 & 45.89220975 & 0.1432 & 11.17 & $1.70$ & 17.16 & 0.999 \\
587725489990402472 & 258.95204855 & 54.28300312 & 0.2154 & 11.47 & $-0.38$ & 17.53 & 0.855 \\
587725491063095592 & 259.24312269 & 56.84922519 & 0.2361 & 11.45 & $0.70$ & 17.36 & 0.955 \\
587725552268738844 & 118.81820385 & 45.28471333 & 0.0503 & 11.42 & $0.26$ & 15.36 & 0.790 \\
587725775067152792 & 115.14104706 & 40.56706267 & 0.1523 & 10.96 & $-0.29$ & 17.25 & 0.863 \\
587725981224468549 & 123.58884576 & 49.84016841 & 0.1691 & 11.19 & $-0.08$ & 17.39 & 0.953 \\
587725981227221163 & 130.23160138 & 54.7043356 & 0.1710 & 10.83 & $0.51$ & 17.57 & 0.978 \\
587725981763371221 & 127.74461264 & 53.85535064 & 0.0631 & 10.98 & $-0.61$ & 15.70 & 0.909 \\
\hline
\end{tabular}
\end{center}
\caption{The first 10 objects in the hybrid visual-CNN CFIS post-merger catalog. The entire post-merger catalog is available digitally via MNRAS.}
\label{pm-table}
\end{table*}

After classification attempts by both the CNN and the authors, galaxies in this study belong to one of three categories:
\begin{itemize}
  \item galaxies given positive (p(x)>0.75) post-merger classifications by the CNN and positive classifications by the authors (699),
  \item galaxies given positive post-merger classifications by the CNN and negative classifications by the authors (1,301), and
  \item galaxies given post-merger classifications p(x)<0.75 by the CNN and which were not inspected visually (166,597).
\end{itemize}

We can also define a more useful sub-category of the third group, for galaxies given classifications of p(x)<0.1 by the CNN, of which there are 131,168. Because the CNN is well-calibrated and was trained on equally sized samples of post-mergers and non-post-mergers, classifications p(x)<0.1 can be confidently interpreted as negative, particularly due to the natural rarity of post-mergers.

\begin{figure*}
\includegraphics[width=\textwidth]{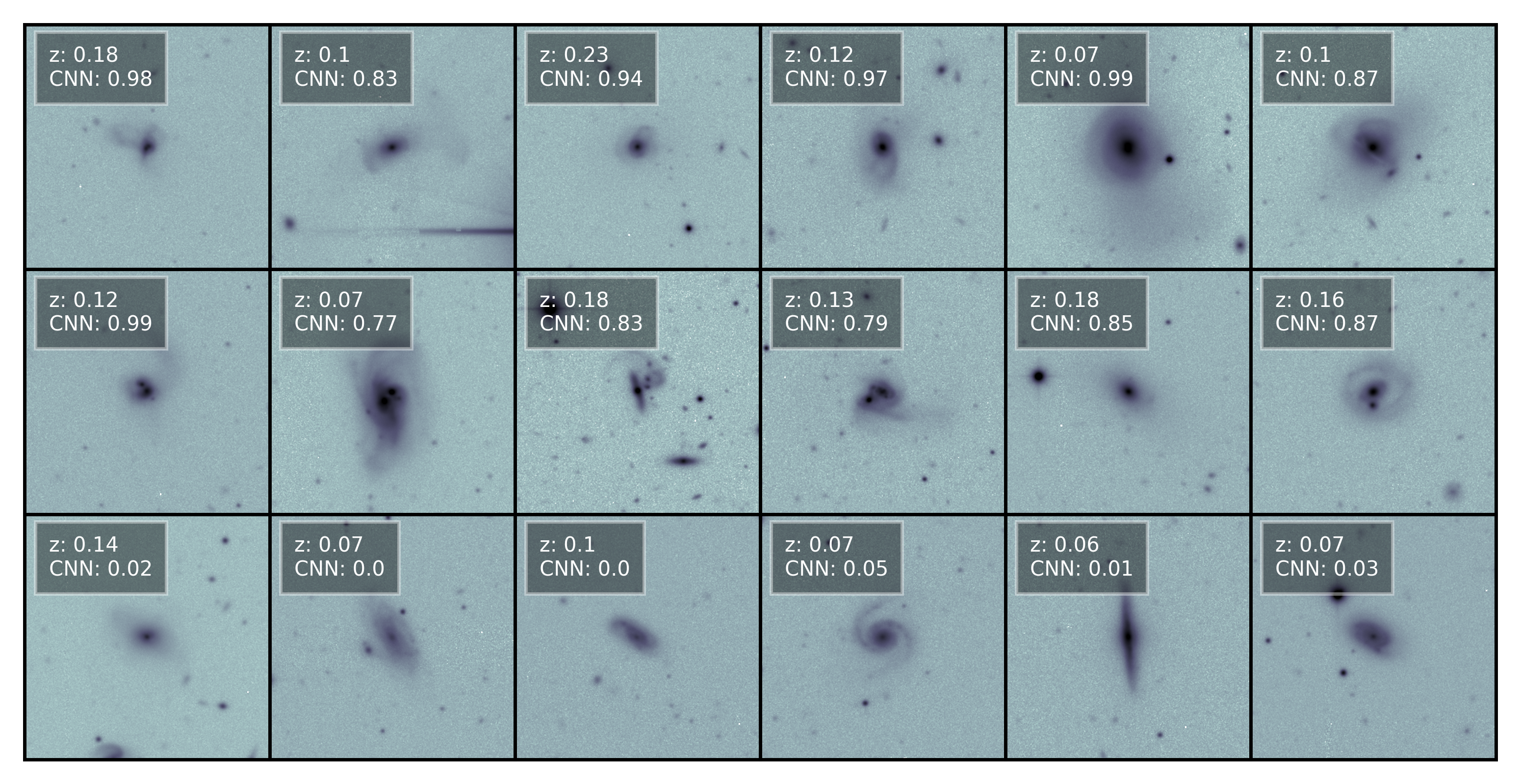}
\caption{CFIS galaxies belonging to each of three categories relevant to this experiment. Top row: a selection of galaxies that received positive (p(x)>0.75) post-merger predictions by the CNN, and which were confirmed as post-mergers by the authors. Merger features in certain galaxies are sometimes more visible upon closer inspection in CFIS imaging. Middle row: Galaxies which received positive post-merger predictions by the CNN which were later ruled out upon close inspection by the authors. Many of these are rejected due to a double nucleus or close companion. Bottom row: Galaxies which received negative post-merger predictions (p(x)<0.1) by the CNN, and which were not inspected visually as part of the experiment. These galaxies are eligible to be used as controls in our subsequent study of $\Delta$SFR in the post-merger epoch.}
\label{fig:agree_mosaic}
\end{figure*}

Images of sample galaxies from the three main categories are shown in Figure~\ref{fig:agree_mosaic}. While varying contrast levels are useful in confirming the presence of merger signatures, the confirmed post-mergers in the top row all have features that are visible with consistent, uniform scaling. In the middle row (rejected galaxies that were predicted to be post-mergers by the CNN), some of the features that deluded the CNN are visible. Notably, while several systems in this category appear to be interacting or pre-coalescence, it is impossible for us to determine whether these galaxies are truly post-mergers which happen to be experiencing a second interaction. Galaxies that received negative (p(x)<0.1, bottom row) predictions tend to have morphologies consistent with settled disks or undisturbed ellipses. For the remainder of this work, star forming post-merger galaxies are taken from the first category, and star forming non-post-merger galaxies (which may be isolated or interacting, but not likely coalesced) are taken from a subset of the CNN-negative pool.

\section{Star formation in post-mergers}
\label{Star formation in post-mergers}

\subsection{Modulating the decision threshold}
\label{Modulating the decision threshold}

Having appraised the CNN's utility as a post-merger identification tool, we can combine its classifications with the authors' visual labels and spectroscopic data from SDSS to address the question of star formation enhancement in post-merger galaxies. We are interested in studying star formation rates of post-merger galaxies relative to other star forming galaxies. As such, it is important that we have robust spectroscopic criteria to identify star forming galaxies, and adequate signal to take reliable measurements of their star formation rates (SFRs). To this end, we use the BPT diagram cut for star forming galaxies described in \citet{2003MNRAS.346.1055K}, which limits the search to galaxies whose emission lines are dominated by star formation. We require a minimum signal to noise ratio of 3 for the lines used in BPT classification (H$\alpha$, H$\beta$, [NII]6584, [OIII]5007). Since the visual classification exercise suggested that the CNN's classifications are unreliable outside of its mass training regime ($\mathrm{10^{10}-10^{12}}$ \(\textup{M}_\odot\), see Figure~\ref{fig:thresh-v-visagree}), we also limit the $\Delta$SFR study to this range. In the resulting population of star forming galaxies, we use total SFRs (extrapolated from the observed SFR from the SDSS fibre spectrum) derived following the methodology of \citet{2004MNRAS.351.1151B}. There are 19,113 galaxies in the final pool eligible for this study (see teal histogram in Figure~\ref{fig:thresh-hist}).

From this work (see Figure~\ref{fig:thresh-v-visagree}, as well as the results of \citealp{2021MNRAS.504..372B}), we know that the CNN's prediction p(x), while imperfect, is strongly linked to post-merger status in both simulated (IllustrisTNG100-1) galaxies, and real galaxies that appear in CFIS. Further, \citet{2021MNRAS.504..372B} performed a simulated $\Delta$SFR experiment in a mock survey of simulated galaxies, and found that a high decision threshold (i.e. the chosen cut in p(x) for a galaxy to be counted as a post-merger) could be effectively used to enforce high post-merger purity in the predicted sample. With a high-purity sample in hand, it is possible to closely estimate the true $\Delta$SFR in recent post-mergers. We take our cue from the mock survey results and investigate $\Delta$SFR with varying p(x) cuts in order to approximate the true enhancement in CFIS post-mergers, albeit at the expense of completeness.

We begin with a set of post-merger decision thresholds, and calculate a median $\Delta$SFR for the galaxies above each. In order to determine the statistical star formation rate enhancement, each post-merger is matched to a set of controls. In order to ensure that the controls are not post-mergers, they are required to have p(x)<0.1. While somewhat arbitrarily chosen, 81\% of galaxies given negative (p(x)<0.5) classifications fall below p(x)=0.1, and the specific choice of non-post-merger control pool does not affect the calculated $\Delta$SFRs. Controls are required to match the post-merger in both stellar mass and redshift, within tolerances of 0.1 dex and 0.01, respectively. If 5 or more eligible controls are found, a $\Delta$SFR is calculated for the predicted post-merger by subtracting the median control SFR from that of the post-merger. If under 5 controls are found, both tolerances are iteratively increased by a factor of 1.5 until at least five are identified. Due to the abundance of controls, 99\% of the post-mergers find five controls without increasing their tolerances, and the remaining galaxies find their controls within 3 growths. Since environmental statistics are not available for all post-mergers, we do not match controls on environmental factors (neighbour distance(s) or neighbourhood density) in order to include as many post-mergers as possible. When environmental factors are folded in, there is a minor suppression ($\leq$0.05 dex) in the calculated $\Delta$SFRs, but all trends remain the same.

\begin{figure*}
\includegraphics[width=\textwidth]{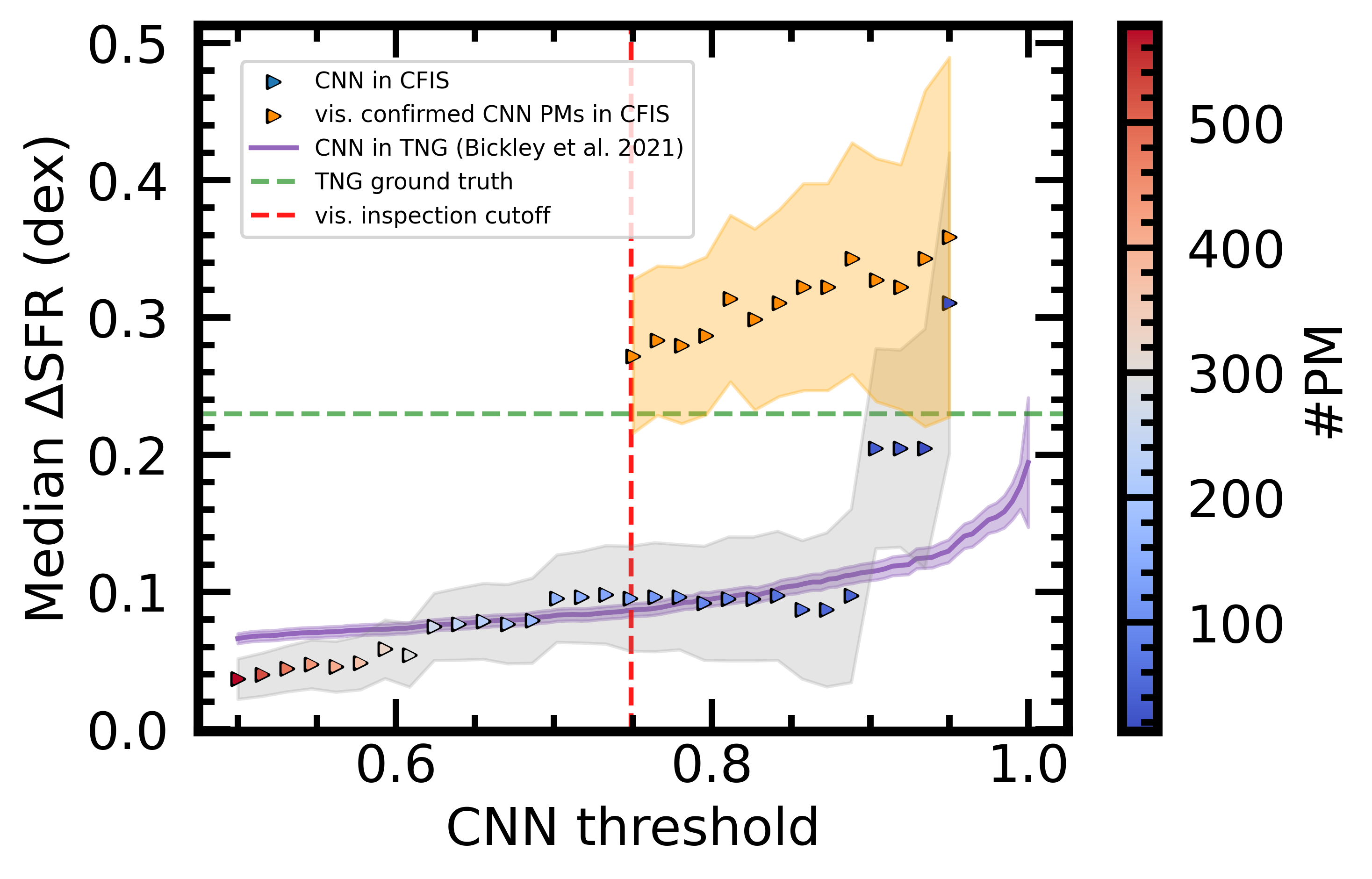}
\caption{Median star formation enhancements found using a range of CNN thresholds. For each data point, all of the galaxies with p(x) higher than the specified value are included in the sample. Error regions represent the statistical error on the median. We compare the trend of increasing star formation with decision threshold (and by extension, sample purity) using 1) only CNN labels (colour-coded points), and 2) the authors' visual labels (orange points) to the trend found by applying the same methodology to galaxies from IllustrisTNG100-1 as in \citet{2021MNRAS.504..372B} (purple curve), as well as the underlying true value for recent post-mergers from IllustrisTNG100-1 (green dashed line). The red dashed line marks the CNN threshold below which galaxies were not inspected visually. Using the CNN alone, we calculate meager $\Delta$SFR for p(x) cuts below 0.6. As in the mock survey, the median $\Delta$SFR increases when more strict p(x) cuts are imposed, up to a maximum of \textasciitilde0.3 dex (or a factor of two) for p(x)>0.95. The visually confirmed sample is already highly enhanced (\textasciitilde0.27 dex) before any additional p(x) cut is applied. As p(x) increases, the median enhancement of the visually confirmed sample increases up to \textasciitilde0.36 dex, though in a less dramatic fashion than the CNN-only sample. The relative stability of the orange series suggests that the visually confirmed sample is already highly pure.}
\label{fig:thresh-v-med-dsfr}
\end{figure*}

Figure~\ref{fig:thresh-v-med-dsfr} shows the results of our star formation enhancement study on predicted post-merger galaxies in CFIS. The green dashed line, included for context, is the true $\Delta$SFR of galaxies in their first post-merger simulation snapshot (<150 Myr since a major merger) in IllustrisTNG100-1 as shown in \citet{2021MNRAS.504..372B}.  The purple curve from the same work shows how the median predicted enhancement evolves as one modulates the decision threshold in a mock survey of simulation galaxies.

The colour-coded points with the grey error region in Figure~\ref{fig:thresh-v-med-dsfr} show the prediction of $\Delta$SFR for CFIS post-mergers made by the CNN alone. As in the simulation, we expect that the CNN identifies increasingly pure CFIS post-merger samples as we enforce increasingly high decision thresholds. As a result, the median $\Delta$SFR of the predicted sample steadily increases between 0.5 < p(x) < 0.9, and makes a more dramatic jump in $\Delta$SFR for the highest decision thresholds. The number of galaxies included naturally decreases with increasing p(x) cuts as well. In Table\ref{thresh-table} we give selected example statistics for varying p(x) cuts, tabulating the number of post-mergers and the measured delta SFR. Although few galaxies are included in the post-merger sample above p(x)=0.9, the IllustrisTNG100-1 results suggest that these data points indicate most accurately the true behaviour of galaxies that have undergone a major merger very recently, with enhancements around 0.3 dex (see the purple curve approaching the green line, Figure~\ref{fig:thresh-v-med-dsfr}).

\begin{table}
\begin{center}
\begin{tabular}{ |c|c|c|c|c| } 
\hline
CNN p(x) & \#PM, full & $\Delta$SFR, full (dex) & \#PM, vis. & $\Delta$SFR, vis. (dex) \\
\hline
\hline
0.5 & 578 & 0.04 & 45 & 0.27 \\
0.8 & 94 & 0.08 & 27 & 0.31 \\
0.9 & 33 & 0.10 & 24 & 0.32 \\
0.95 & 13 & 0.31 & 7 & 0.36 \\
\hline
\end{tabular}
\end{center}
\caption{The number of CNN-predicted and visually confirmed post-mergers with CNN predictions above selected p(x) cuts, with Median $\Delta$SFR values for each subset. See also Figure~\ref{fig:thresh-v-med-dsfr}.}
\label{thresh-table}
\end{table}

The orange series in Figure~\ref{fig:thresh-v-med-dsfr} shows the result of the same experiment for the visually confirmed post-merger sample. Without any additional cuts (other than the initial p(x)>0.75 used to select galaxies for visual inspection), the visually confirmed galaxies are already highly enhanced in star formation relative to controls, producing nearly twice as much stellar material on average than their non-post-merger counterparts. Applying a progressive decision threshold to the visual sample does not produce a dramatic spike in enhancement for high p(x) as it does in the CNN-selected samples, suggesting that the visual classification exercise has already produced a highly pure sample, but it does further enhance the median predicted $\Delta$SFR up to \textasciitilde0.36 dex.

\begin{figure}
\includegraphics[width=\columnwidth]{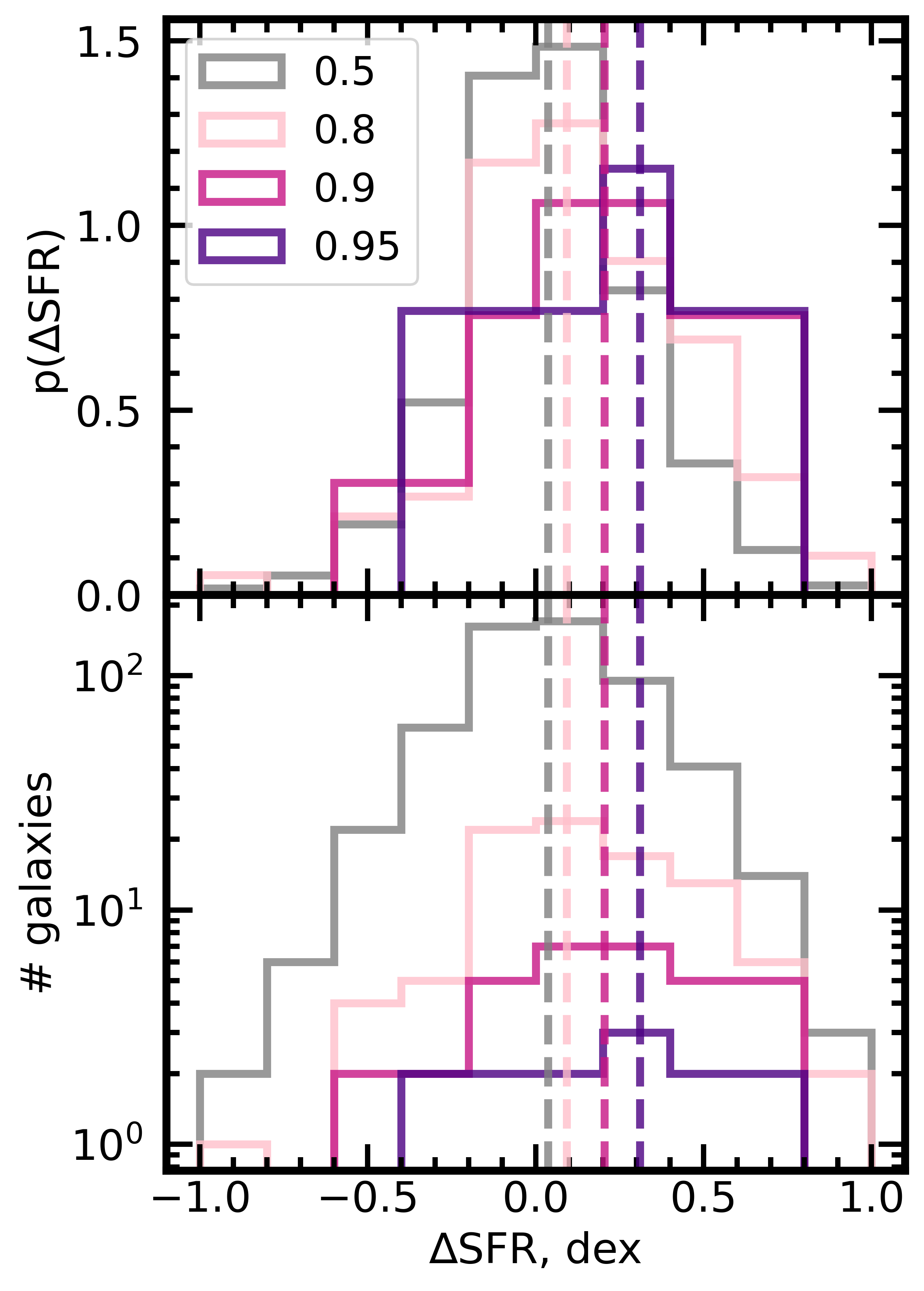}
\caption{The normalized (top panel) and log-scale histogram (bottom panel) distributions of the star formation enhancements calculated for selected CNN thresholds, with vertical dashed lines representing the median enhancements for each group. As the threshold grows more extreme, from p(x)>0.5 (grey curve) to p(x)>0.8 (pink curve), p(x)>0.9 (violet) and p(x)>0.95 (dark purple), the sub-samples of galaxies are increasingly enhanced on average.}
\label{fig:sfr-hist-prog}
\end{figure}

Figure~\ref{fig:sfr-hist-prog} shows the distribution of $\Delta$SFR for CNN-identified post-mergers at selected p(x) decision thresholds: p(x)>0.5, the natural threshold for a well-calibrated model trained on balanced data, p(x)>0.8, p(x)>0.9, and p(x)>0.95. The normalized (top) version of the figure shows the change in the shape of the distribution of post-merger $\Delta$SFRs, while the non-normalized logscale version of the figure (bottom) shows how the successively smaller (and presumably more pure) subgroups of predicted post-mergers move towards a higher median $\Delta$SFR as shown in Figure~\ref{fig:thresh-v-med-dsfr}. It is noteworthy that for a sample with p(x)>0.5, we find a negligible median enhancement in the SFR of post-mergers thus selected. An effect analogous to the null result found when using a cut at p(x)=0.5 is likely responsible for the under-prediction of merger induced star formation enhancement in recent literature (e.g. \citealp{2019A&A...626A..49P}).

The trends uncovered with the CNN alone in Figures~\ref{fig:thresh-v-med-dsfr} and~\ref{fig:sfr-hist-prog} could be explained by an intuitive physical connection between merger intensity, the time since a galaxy underwent a merger, the prominence of visible merger-induced morphologies, and the strength of merger-induced star formation enhancement. To take an extreme example, we might observe a galaxy which has just coalesced after an equal-mass major merger just millions of years ago. Such an object would have prominent stellar streams and/or shells, which would be obvious to the neural network. Such a galaxy might be assigned a p(x) of 0.95 by the CNN, and found to have a large $\Delta$SFR. However, merger status is not truly binary: galaxies may be identified as post-mergers by the CNN, but exhibit less extreme morphology and weaker enhancement, and galaxies which experienced minor mergers Gyr ago are very likely to go undetected. In interpreting our results, it is useful to imagine a continuum of merger statuses between these hypothetical extremes. Still, we can report post-merger star formation enhancements with some confidence: \textasciitilde0.3 dex for an inclusive, visually confirmed sample of post-mergers.

\subsection{Star formation in the merger sequence}
\label{Star formation in the merger sequence}

\citet{2013MNRAS.435.3627E}, showed how observed galaxy characteristics from the SDSS correlate with pair phase separation, and traced these trends into the post-merger epoch via visual classifications made by the authors. For many of these properties, including AGN excess, fractional composition of young stars, star-forming galaxy fraction, and $\Delta$SFR, they found the post-merger epoch to be the most transformative. In general, after galaxies undergo their first pericentric passage, the galaxies begin to influence one another. As the dynamics of the participant galaxies become increasingly disrupted, gas is funneled into the centres, where the increased gas surface density leads to new star formation and AGN activity. By the time the merging galaxies have fully coalesced into a single, self-governing entity, the pace of these processes reaches its highest intensity. \citet{2012MNRAS.426..549S} and \citet{2013MNRAS.433L..59P} also traced these correlations out to separations as wide as 150 kpc, and made a strong case for merger-induced activity by comparing SDSS pair observations to theoretical models.

We seek to compare our findings in CFIS to the result found by \citet{2013MNRAS.435.3627E} for $\Delta$SFR by calculating a typical enhancement for post-merger galaxies using our new, larger, machine-identified sample. We also adjust the control matching procedure from \citet{2013MNRAS.435.3627E} in order to maintain consistency with the method used to match IllustrisTNG post-mergers and controls in \citet{2021MNRAS.504..372B} and earlier in this work. Using projected pair separation statistics from \citet{2016MNRAS.461.2589P}, we measure median $\Delta$SFR values in bins of projected separation (in kpc) for galaxy pairs in much the same way as we do for post-mergers. We match star forming (defined the same as for post-mergers) galaxy pairs with separations of < 80 kpc to controls, which must have separations of > 80 kpc and Galaxy Zoo (\citealp{2010MNRAS.401.1043D}) merger vote fractions of zero, on redshift and stellar mass. The default matching tolerances and growth methodology is the same as for post-mergers.

\citet{2013MNRAS.435.3627E} use 97 visually identified galaxies. Applying our $\Delta$SFR methodology (which is similar to that used by Ellison et al. 2013) yields a sample of only 26 galaxies, highlighting the small sample statistics of previous work. For comparison with the Ellison et al. (2013) post-merger sample, we select two sub-samples of CFIS mergers - one found using the CNN with a decision threshold at 0.8, and another for the entire visually confirmed post-merger sample. There are 94 and 45 galaxies in these two sub-samples, respectively.

\begin{figure}
\includegraphics[width=\columnwidth]{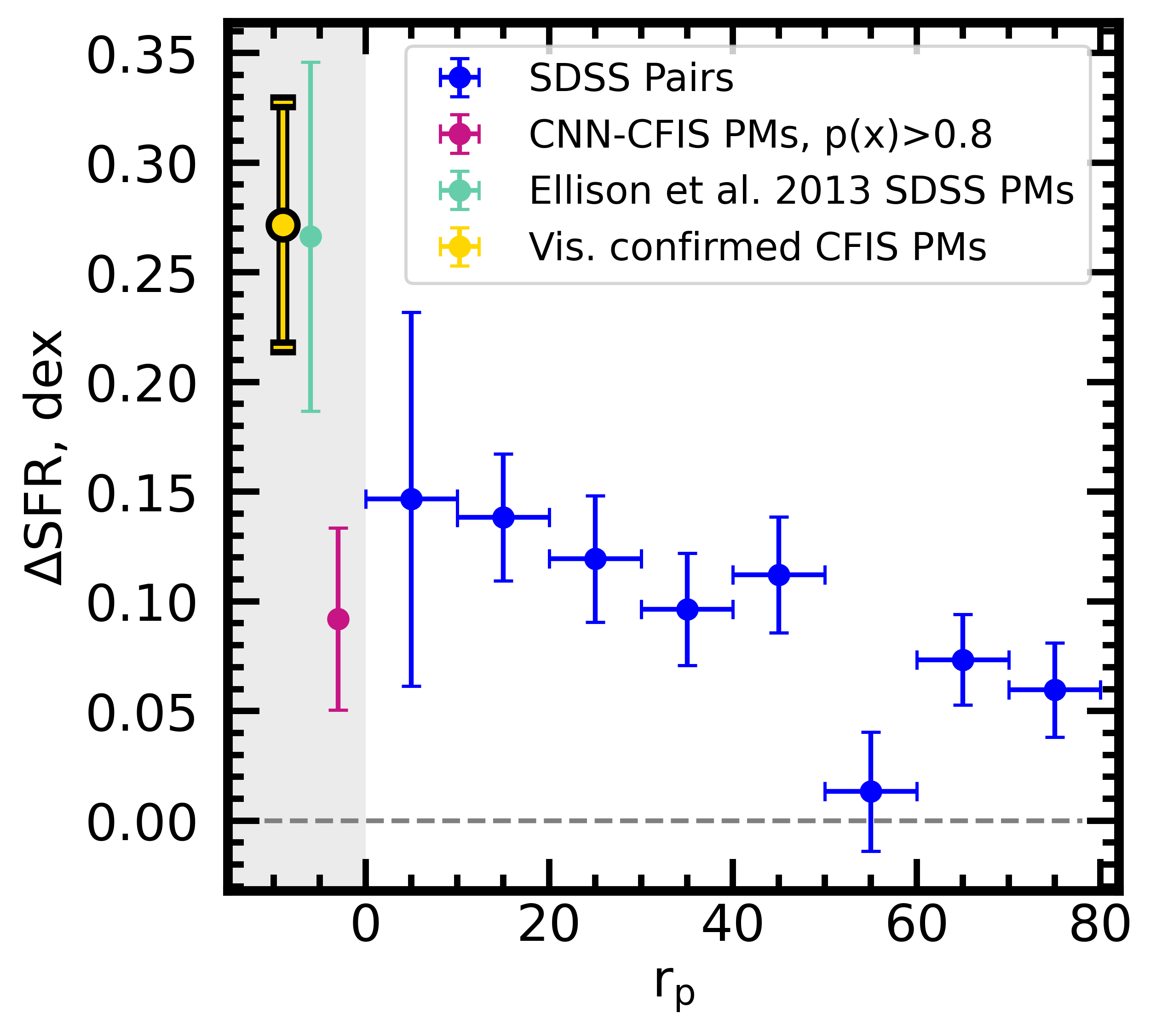}
\caption{The trend between projected galaxy pair separation in SDSS-DR7 and star formation enhancement (blue series) in the style of \citet{2013MNRAS.435.3627E} using matching, star formation criteria, and signal to noise criteria consistent with this work. We uncover the same trend; that $\Delta$SFR increases as a function of projected separation as repeated pre-coalescence interactions funnel gas reserves towards the centres of the participant galaxies. Of the \textasciitilde100 SDSS post-mergers in \citet{2013MNRAS.435.3627E}, 26 survive our selection criteria (turquoise point). They are even more highly enhanced in star formation than the closest pairs. Without the assistance of visual labels, our CNN uncovers increasingly high median enhancements as a function of decision threshold (see Figure~\ref{fig:thresh-v-med-dsfr}) in increasingly small post-merger samples. Of our visually cleaned sample, 45 galaxies survive our selection criteria. They are as strongly enhanced on average as the post-mergers from \citet{2013MNRAS.435.3627E}.}
\label{fig:rp-v-dsfr}
\end{figure}

Figure~\ref{fig:rp-v-dsfr} shows that we uncover a trend consistent with \citet{2013MNRAS.435.3627E}  in the pair phase, with a peak $\Delta$SFR of \textasciitilde0.15 dex for pairs with projected separations < 10 kpc. Without the help of visual inspection, even with a relatively high threshold of p(x)>0.8, the CNN predicts a median $\Delta$SFR of 0.08 dex for the post-mergers, which is lower by \textasciitilde0.07 dex than for close pairs with projected separations below 10 kpc. However, the median enhancement for the entire population of visually confirmed post-mergers meeting the criteria for the study is consistent with the \citet{2013MNRAS.435.3627E} result at \textasciitilde0.27 dex, confirming the link between the post-merger epoch and strongly enhanced star formation in a sample that is nearly twice as large. Further, this result is consistent with recent theoretical studies of IllustrisTNG by \citet{2020MNRAS.494.4969P} and \citet{2020MNRAS.493.3716H}, suggesting that IllustrisTNG served as a suitable training ground for the CNN.

\section{Summary}
\label{Summary}

We have applied a convolutional neural network (CNN) trained to identify simulated, realism-added, recently-coalesced post-mergers from the IllustrisTNG simulations to $r$-band galaxy images from the deep and high-resolution Canada-France Imaging Survey (CFIS). The results of this application and the subsequent study of the out-falling predicted post-merger galaxies are summarized here:

\begin{itemize}
  \item The global statistics of the merger status predictions (given by the CNN on a sliding scale of 0 to 1 roughly analogous with post-merger certainty) are consistent with our expectations given the rarity of post-mergers in the Universe (Figure~\ref{fig:thresh-hist}), and considering the results of a proof-of-concept "mock survey" detailed in \citet{2021MNRAS.504..372B}.
  \item Visual inspection of 2,000 galaxies which received CNN post-merger predictions p(x)>0.75 shed light on the network's strengths and shortcomings (Figures~\ref{fig:confirmed-stats} - \ref{fig:agree_mosaic}). Only about one third (699) of the galaxies with CNN predictions p(x)>0.75 were confirmed as post-mergers upon visual inspection by the authors, and the network is qualitatively successful in identifying post-merger-like features in CFIS images. When the network errs, it is most commonly due to a barely-separated double nucleus. These double nuclei typically required close inspection on a high-contrast setting to identify by eye. Less frequently, the network was confused by a more visually obvious interacting companion in the CFIS cutout or wider SDSS imaging, or a survey artifact (chip gap, saturated foreground object, or incomplete survey segment).
  \item We study the star formation enhancements of post-mergers identified by the CNN, as well as those confirmed by eye. We find that using a decision threshold, i.e. a cut in CNN merger status prediction, appears to enhance the purity of the CNN-predicted post-merger sample in CFIS. As the decision threshold and sample purity increase, we find that the median $\Delta$SFR increases as well. For the highest decision thresholds p(x)>0.95, $\Delta$SFR peaks at \textasciitilde0.3 dex (Figures~\ref{fig:thresh-v-med-dsfr} and~\ref{fig:sfr-hist-prog}). This increasing behaviour is qualitatively consistent with the mock survey result from \citet{2021MNRAS.504..372B}.
  \item The $\Delta$SFR values for the visually confirmed post-mergers are significantly higher than the CNN-only population for all decision thresholds, and exceeds the peak value of the uninspected post-merger sample (0.36 dex on average) when a high decision threshold cut is enforced (Figure~\ref{fig:thresh-v-med-dsfr}). Together, all the visually confirmed post-mergers that meet the spectroscopic criteria for the $\Delta$SFR experiment have a median enhancement of \textasciitilde0.27 dex.
  \item We revisit one of the main results of \citet{2013MNRAS.435.3627E}, the trend in $\Delta$SFR with merger stage. In qualitative agreement with \citet{2013MNRAS.435.3627E}, we find that pair-phase galaxies are increasingly enhanced in star formation as they grow closer to their merging companions. The improved depth of CFIS over SDSS and the efficiency of the CNN as a sample distillation tool allow us to revisit the 2013 result in a post-merger sample that is nearly twice as large. The sample of visually confirmed post-mergers in this work are as enhanced as the post-mergers identified by \citet{2013MNRAS.435.3627E} using only SDSS imaging and visual inspection (Figure~\ref{fig:rp-v-dsfr}).
\end{itemize}

The impact of our new visually confirmed post-merger sample extends beyond this work. Observations and theory both predict that post-mergers ought to be significantly enhanced in AGN activity as well as star formation. Moreover, the impact of a recent merger on AGN luminosity may be longer-lived and more easily detectable than enhanced star formation. In a future publication, we will therefore study the statistical connections between merger status as predicted by the CNN and AGN activity, again using SDSS optical spectra to characterize our CFIS post-mergers.

\appendix
\section{Brightness, mass, and redshift}
\label{appA}

\begin{figure}
\includegraphics[width=\columnwidth]{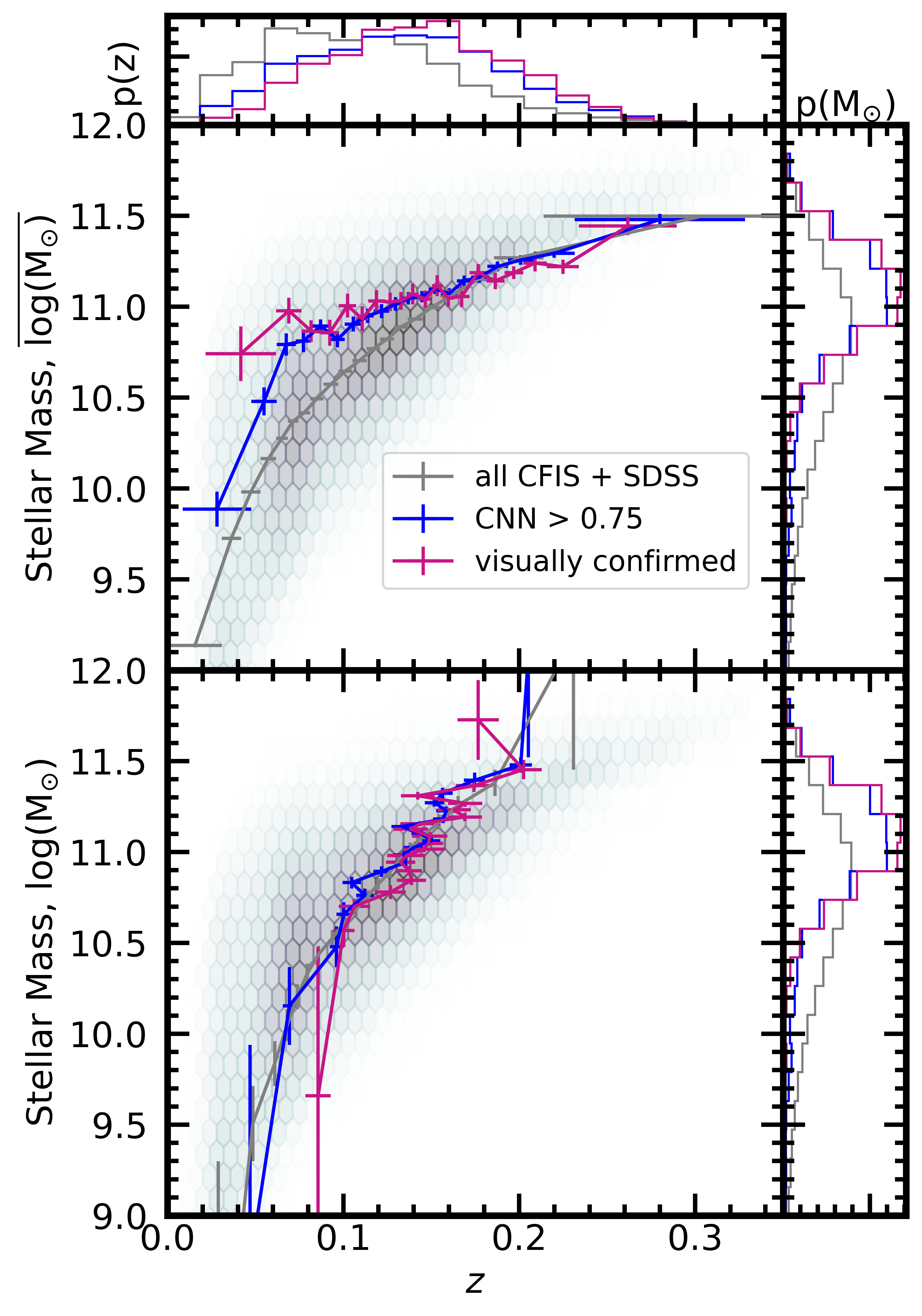}
\caption{The average stellar mass in equal-numbered bins of redshift (top panel) and the average redshift in equal-numbered bins of stellar mass (bottom panel) for the parent sample of CFIS galaxies with SDSS spectra (grey curve), the sample of galaxies assigned predictions over 0.75 by the CNN (blue), and the visually confirmed sample (violet). Both panels are superimposed over density histograms of the parent sample in the background. 1D histograms for each statistic are shown in the auxiliary panels. The presence of a redshift-dependent mass bias and the lack of a mass-dependent redshift bias together suggest that the relative brightness of a galaxy in its respective CFIS image may bear on its likelihood to be selected by the CNN. This preference by the CNN is likely responsible for some unusual statistical characteristics of the visually confirmed post-merger sample.}
\label{fig:bias-app}
\end{figure}

In Section~\ref{The hybrid classification post-merger sample}, we present the demographics of the CNN-identified and hybrid post-merger galaxy samples and compare them to the parent sample of CFIS galaxies with SDSS spectra. We posit that the neural network's selection function is sensitive to the brightness of a galaxy, and by extension, the brightness of any tidal features, relative to the image background. The top panel of Figure~\ref{fig:bias-app} shows the median stellar mass in adaptive bins (i.e. the same number of galaxies fall into each bin) of redshift, while the bottom panel shows the median redshift in adaptive bins of stellar mass. Both panels have a hexagonal 2D histogram of the parent sample in the background. The median statistic for the parent sample is shown in grey for both panels, the CNN-identified subset is shown in blue, and the visually confirmed galaxies are shown in violet. The auxiliary panels show density histograms for each quantity in the same colour scheme.

The top panel of Figure~\ref{fig:bias-app} shows that in a given redshift bin, the average stellar mass is higher for the CNN and visually confirmed samples than in the parent sample. The network is therefore more likely to select a higher-mass galaxy over a lower-mass galaxy if they both appear at the same redshift. This is a natural consequence of the fact that more massive galaxies are more likely to be bright relative to the background at a given luminosity distance, and their tidal features are more likely to stand out (both visually and numerically). This effect weakens somewhat at higher redshift, where galaxies in the parent sample are more massive on average, and the role of distance in determining the imaged brightness of a galaxy may become more significant.

The bottom panel (viewed with the vertical axis as the independent variable) shows that the inverse trend is not true. In all stellar mass bins, the median redshifts of the parent, CNN, and visually inspected samples are in good agreement. Put another way, if two equal-mass galaxies appear at different redshifts in the survey, the network is no more likely to select one than the other. As such, the statistical biases of the sample in mass and redshift can both be comfortably interpreted as a result of each image's signal (galaxy brightness) to noise (sky background, seeing, and any survey artifacts) ratio.

\section*{Acknowledgements}
\label{Acknowledgements}

The work detailed above was conducted at the University of Victoria in Victoria, British Columbia, as well as in the Township of Esquimalt in Greater Victoria. We acknowledge with respect the Lekwungen peoples on whose unceded traditional territory the university stands, and the Songhees, Esquimalt and $\mathrm{\ubar{\mathrm{W}}S\acute{A}NE\acute{C}}$ peoples whose historical relationships with the land continue to this day.

CFIS is conducted at the Canada-France-Hawaii Telescope on Maunakea in Hawaii. We also recognize and acknowledge with respect the cultural importance of the summit of Maunakea to a broad cross section of the Native Hawaiian community.

This work is based on data obtained as part of the Canada-France Imaging Survey, a CFHT large program of the National Research Council of Canada and the French Centre National de la Recherche Scientifique, and on observations obtained with MegaPrime/MegaCam, a joint project of CFHT and CEA Saclay, at the Canada-France-Hawaii Telescope (CFHT) which is operated by the National Research Council (NRC) of Canada, the Institut National des Science de l’Univers (INSU) of the Centre National de la Recherche Scientifique (CNRS) of France, and the University of Hawaii. This research used the facilities of the Canadian Astronomy Data Centre operated by the National Research Council of Canada with the support of the Canadian Space Agency.

We thank Florence Durret and Martin Kilbinger for their insightful comments on this paper.

Data from the IllustrisTNG simulations are integral to this work. We thank the Illustris Collaboration for making these data available to the public.

Funding for the SDSS and SDSS-II has been provided by the Alfred P. Sloan Foundation, the Participating Institutions, the National Science Foundation, the U.S. Department of Energy, the National Aeronautics and Space Administration, the Japanese Monbukagakusho, the Max Planck Society, and the Higher Education Funding Council for England. The SDSS Web Site is http://www.sdss.org/. The SDSS is managed by the Astrophysical Research Consortium for the Participating Institutions. The Participating Institutions are the American Museum of Natural History, Astrophysical Institute Potsdam, University of Basel, University of Cambridge, Case Western Reserve University, University of Chicago, Drexel University, Fermilab, the Institute for Advanced Study, the Japan Participation Group, Johns Hopkins University, the Joint Institute for Nuclear Astrophysics, the Kavli Institute for Particle Astrophysics and Cosmology, the Korean Scientist Group, the Chinese Academy of Sciences (LAMOST), Los Alamos National Laboratory, the Max-Planck-Institute for Astronomy (MPIA), the Max-Planck-Institute for Astrophysics (MPA), New Mexico State University, Ohio State University, University of Pittsburgh, University of Portsmouth, Princeton University, the United States Naval Observatory, and the University of Washington.

This research was enabled, in part, by the computing resources provided by Compute Canada.

\section*{Data Availability}
\label{Data Availability}

Simulation data from TNG100-1 used in the generation of training images for this work are openly available on the IllustrisTNG website, at tng-project.org/data. Template versions of \textsc{RealSim} and \textsc{RealSim-CFIS}, developed by Connor Bottrell with modifications by RWB are publicly available via GitHub at github.com/cbottrell/RealSim and github.com/cbottrell/RealSim-CFIS. Specific image training data used to develop the findings of this study are available by request from RWB.

The visually confirmed post-merger catalog is included for download with the MNRAS release of this article.



\bibliographystyle{mnras}
\bibliography{bib_01} 






\bsp	
\label{lastpage}
\end{document}